\documentclass[12pt,preprint]{aastex}
\slugcomment{To Appear in the Astrophysical Journal}

\shorttitle{Hard X-ray Emission from the NGC 5044 Group}
\shortauthors{Mark Henriksen}

\begin{document}

\title{Hard X-ray Emission from the NGC 5044 Group}

\author{Mark J. Henriksen}
\affil{Physics Department, University of Maryland, Baltimore County
    Baltimore, MD 21250}

\begin{abstract}

Observations made with the Rossi X-ray Timing Explorer (RXTE)  Proportional Counter Array (PCA) to constrain the hard X-ray emission in the NGC 5044 group are reported here. Modeling a combined PCA and ROSAT position sensitive proportional counter (PSPC) spectrum with a 0.5 - 15 keV energy range shows excess hard emission above 4 keV. Addition of a  powerlaw component  
with spectral index of 2.6 - 2.8 and 
luminosity of 2.6$\times$10$^{42}$ ergs s$^{-1}$ within 700 kpc in the observed energy band removes these residuals.
Thus, there is a detection of  a significant non-thermal component that is 32\% of the total
X-ray emission. 
%Several bright discrete sources are visible in the PSPC detector region that encompasses the RXTE field of view. 
%The emission from these point sources is combined to simulate the contribution of point
%sources to the PCA spectrum. 
%The co-added spectrum is best fit by
%a two component model consisting of a thermal model with 0.93 - 1.13 keV and luminosity of
%1.45$\times$10$^{41}$ ergs s$^{-1}$ plus a powerlaw, $\alpha$ = 2.08 - 4.12, and luminosity 4.48$\times$10$^{41}$ ergs s$^{-1}$. 
Point source emission makes up at most 14\% of the non-thermal emission from the NGC 5044 group. Therefore, the diffuse, point source subtracted, non-thermal component is 2.2 - 3.0$\times$10$^{42}$ergs s$^{-1}$.
% In addition, the X-ray derived redshift for each of the models models is at least 0.045 with 90\% confidence,
%higher than the optical redshift, 0.01. 
%Assuming the non-thermal and thermal pressure equilibrium at the position of
%the lobe results in a k parameter, the ration of cosmic-ray proton to cosmic-ray electron energy density, of 100. 
The cosmic-ray electron 
energy density is 3.6$\times$10$^{-12}$ ergs cm$^{-3}$ and the average magnetic field is 0.034 $\mu$Gauss in the largest  radio emitting region. The ratio of
cosmic-ray electron energy density to magnetic field energy density, $\sim$2.5$\times$10$^{4}$,  is significantly out of equipartition and is therefore atypical
of radio lobes. In addition, the group's small size and low non-thermal energy density 
strongly contradicts the size-energy relationship found for radio lobes. Thus, it is unlikely to be related to the active galaxy and is most likely a relic of
the merger. The energy in cosmic-rays and magnetic field is consistent with simulations of cosmic-ray acceleration by
merger shocks.

\end{abstract}


\section{Introduction}

Many clusters and groups of galaxies have diffuse regions of radio emission typically characterized as halos, relics,  or lobes. Evidence of non-thermal X-ray emission in clusters (references in Rephaeli et al. 2008) and groups (Nakazawa et al. 2007; Fukazawa et al. 2001; Hudson \& Henriksen 2003) has also been found. One source of the cosmic-rays that provide the radio and X-ray emission is
accretion shocks (Kushnir, Katz \& Waxman 2009).  Approximately 5 - 60\%
of the accretion shock energy is expected to go into cosmic-ray production (Kang, Jones \& Gieseler 2002) producing 
the non-thermal emission. 
Simulations predict that accretion shocks will produce nonthermal emission in 
a broad mass range $\sim$10$^{13}$ - 10$^{14}$ (Miniati et al. 2001) of mergers, which includes not only rich
clusters but groups of galaxies. 

Another source of cosmic rays are Active Galactic Nuclei (AGN) outburst (Fujita et al. 2007).
Hydrodynamical simulations of the intragroup medium show that, without a heating source, cooling in the core leads to compression of gas which predicts, in the X-ray band, higher luminosities than observed. Inclusion of a heating source, for example AGN outflow
driven by accretion onto a supermassive blackhole, reduces the core luminosity to observed levels (Puchwein, Sijacki, \& Springel 2009).
Heating and cooling rates are sufficiently close in clusters to make this scenario plausible (Fabian \& Sanders 2009). The AGN outflow is correlated with the cavities in the intracluster/intragroup  medium which appear as dark holes in the X-ray emission. For the
HCG 62 group, the cavities are in approximate pressure balance with the surrounding thermal medium (Gitti et al. 2009). 
The NGC 5044 group shows X-ray cavities among plumes of gas from a highly asymmetric X-ray core (Gastaldello et al. 2009; David et al. 2009). The latter authors estimate the power from radio quiet cavities, whose
energy is deposited locally, within 10 kpc, to be sufficient to heat about half of the gas. The presence of an H$\alpha$ filament and interstellar
FIR emission (Temi, Brighenti, \& Mathews 2007) suggests that some of the gas is still cooling. David et al. (2009) report
a larger cavity to the south of the core that is radio filled and releasing substantially more energy outside of the core but within 25 kpc.
David et al. (2009), Gastaldello et al. (2009), and Buote et al. (2003) have analyzed the XMM and Chandra observations in detail and present a comprehensive picture of the complex dynamical state  of the NGC 5044 group, which includes cavities, filaments, and a cold front.

%An NVSS radio source is located at (13 15 23.9 +/- 0.04, -16 23 7.6 +/- 0.1) coincident with the NGC 5044 galaxy. The flux at 1.4 GHz is 34.7 +/-1.1 mJy.  The major and minor axes are $<$21.8 and $ <$15.5 arc sec, respectively.
% This indicates that the GHz radio source is less than about 4 kpc and perhaps confined to the core of the galaxy.  
David et al. (2009) present GMRT observations at 235 MHz and 610 MHz that reveal several diffuse
radio sources extending well beyond the central galaxy and out into the intragroup medium, approximately 82 kpc.
%The left panel of figure 2 shows the core X-ray emission obtained
%with Chandra contoured over the DSS red image of the group. The right panel shows the NVSS radio contours over the
%X-ray image of the core.  
The radio sources are evidence of cosmic-rays which may also be sources of X-ray emission via inverse-Compton
scattering of the Cosmic Microwave Background photons.

In this paper, we report on observations of the NGC 5044 group made with RXTE to detect such non-thermal X-ray emission.
RXTE has a large effective area, in excess of 1000 cm$^{2}$ at energies in the range of 5 - 12 keV, making it far superior to Swift, Suzaku, XMM, Chandra, and ASCA
in sensitivity to non-thermal emission from groups and low temperature clusters. The RXTE PCA is
also ideal for detecting non-thermal gas from a low temperature cluster or group because it has high sensitivity up to 15 keV without significant particle
background contamination. Thermal emission is generally predicted to increase more strongly than non-thermal emission with increasing temperature (or mass).  As a result, detection of non-thermal emission by non-imaging instruments is increasingly difficult as contamination
from the thermal component becomes increasingly dominant at higher temperature. The PCA provides a hard X-ray band above that available from 
imaging detectors on Chandra and XMM. This energy band is important for group detection because the
thermal emission from a 1 keV emitter falls off dramatically above 8 keV and the non-thermal emission dominates.
Thus, separation of non-thermal from
thermal emission becomes more reliable. For a steep spectrum source such as radio halo, the 500 - 1000 eV range can also provide
an important constraint on the powerlaw. A detector such as the ROSAT PSPC is ideal to constrain the powerlaw as well as the
thermal emission from group. 
With these observations we are able to detect hard X-ray
emission from AGN or merger shock accelerated cosmic-rays. Under the assumption that the radio and non-thermal X-ray
emission originate from the same population of electrons, we have also constrained
the spectral index of the largest and strongest radio source, to the southeast, that is detached from the central radio source in NGC 5044.
The X-ray and the radio fluxes are used to
calculate the energy density in cosmic rays and magnetic field (without assuming equipartition) to determine the origin of the cosmic-rays.

All luminosities quoted in this paper are for the analyzed energy band, 0.5 - 15 keV, and the analyzed region, 700 kpc (H$_{0}$=73 km s$^{-1}$ Mpc$^{-1}$, $\Omega_{m}$=0.27, $\Omega_{v}$=0.73). 

\section{Observations and Analysis}

The RXTE PCA was used to observe the NGC 5044 galaxy group for 258,400 seconds. Data from the top layer of
PCA arrays 2 and 3 were used to extract a spectrum. Most of the counts are in the top layer
and 2 and 3 were turned on most of the time as 1 and 4 were only used sporadically. Thus, this combination
of detectors will provide the highest signal-to-noise. 
Various filters were applied to the PCA data in constructing
a Good Times Interval (GTI) filter for extracting a spectrum. This includes filtering out bright Earth, 
slews, high electron contamination, and times near South Atlantic Anomaly passage. 
Still, there were spikes visible in the resulting light curve. A count
rate filter was applied to the light curve
to eliminate spikes and obtain a reasonably flat light curve for the extracted spectrum.
The resulting PCA spectrum's energy band is 3.7 - 14.8 keV with a background
subtracted count rate of 0.39 +/- 0.008 counts s$^{-1}$. 

The ROSAT PSPC was used to extend spectral coverage to the soft X-ray regime and provide a strong constraint on non-thermal
as well as thermal emission , which is below the peak sensitivity of the PCA. The PSPC is preferable to other imaging detectors
because it has a wide field of view
that provides full coverage of the 
%The ROSAT X-ray image shows that hot gas extends well beyond the central galaxy, NGC 5044, to a radius of 4 arcmin, or 
%46.5 kpc. 
RXTE PCA field of view, HWHM of 1 degree, or 700 kpc. 
An extraction region of circular shape with radius of 30 arc min was to filter the PSPC data. Visual inspection of the PSPC image
shows that the emission is confined to this radius.
The flux calibration of the PCA, which is important for joint fitting with the PSPC, is discussed at length in Jahoda et. al. (2006) and readers are directed to that publication. We note a couple issues most relevant to the NGC 5044
spectral modeling. The dead time correction is important in matching the absolute flux of the Crab. The estimated dead time is 10$^{-5}$ seconds per event.
For a high count rate object, such as the Crab, 13,000 cps, this is 0.13 dead sec/sec and can lead to an underestimate of
the Crab flux by 6\% if not accounted for. However, for NGC 5044, with a gross PCA count rate of 17.5 cps, has a total dead time of 45 sec over the 260 ksec observation, which is negligible for the absolute flux calibration of the NGC 5044. Additional tweaking of the effective areas to match the Crab flux has been included in the version of {\it xpcaarf} in HEASOFT 6.9 tools, which were used in this analysis. These corrections remove a substantial fraction of the discrepancies noted by Kuulkers et al. (2003) and Revnivtsev et al. (2003)(Jahoda et al. 2006).
Another issue in absolute flux calibration relevant to joint fitting of the spectra involves correcting off axis emission
for the detector response. Using data given in Jahoda et al. (2006) for the PCA collimator response together with the ROSAT radial surface brightness profile parameters, $\beta$ = 0.52, core radius = 48 arc sec (Buote et al. 2003; David et al. 1994) we calculated that a 2.5\% increase in the RXTE source count rate is needed to match the PSPC  within 30 arc min. The PCA source count was increased by this factor prior to spectral modeling. A similar procedure is done in Lutovinov et al. (2008) for the Coma cluster. Spectral fits to the Crab data (Shaposhnikov, Jahoda, \& Markwardt 2009) for the PCA have been improved to a residual 0.5\% systematic error. The improvements in pcarmf include modeling of Xenon L-escape features, a better channel to energy conversion table, and requiring a better fit to the Crab canonical spectrum. Only a systematic error, 0.5\%, that needs to be included in the spectral fitting procedure, which was done here.

The PCA background is well modeled using blank sky observations with a scaling up of the background by an additional 2.5\%. This factor
was found to minimize $\chi^{2}$ compared to scaling factors ranging from 0 to 5\%. Particle background is minimized using the GTI
filter described above. There are residual spatial fluctuations of around 7 - 8\% in the hard CXB due to unresolved point sources that must be modeled.
For the PSPC, the background was obtained in several source free regions in the image. The resulting ROSAT PSPC spectrum is 26,800 seconds in duration with energy band  0.5 - 2.5 keV and background 
subtracted count rate of 3.2 +/- 0.01 counts s$^{-1}$. Excluding data below 0.5 keV eliminates most of the soft X-ray background
while the PSPC response is significantly impaired above 2.5 keV. 

Both spectra were fit jointly creating a 0.5 - 15 keV spectrum with a small gap between
2.5 - 3.7 keV. Three models
were fit to the spectrum: a single thermal, a thermal and a powerlaw, and two thermal components. The hard cosmic X-ray background (CXB) 
fluctuations were modeled in the PCA spectrum. The CXB contribution was modeled as a powerlaw with index -1.29, cutoff at 41.13 keV, and
variable amplitude equal to 8\%  of the mean CXB flux at 20 keV (normalization of +/-1.84$\times$10$^{-4}$) (Revnivtsev et al. 2003). 
We used the Raymond and Smith model in XSPEC for the thermal emission. 
Comparisons between the Raymond and Smith code and APEC, for example, shows generally good agreement for important Fe and O spectral features (Smith et al. 2001) in the PSPC/RXTE bandpass. While the shape of the FeL complex for a 1 keV plasma differs somewhat in the two codes, the  equivalent width appears the same,
which is important since the PSPC only detects integrated emission.
Free parameters for a single thermal component model are  temperature, abundance, and normalization, which is proportional to the emission integral of the extraction region and CXB
normalization. Next, a second component was added, either a power law,
or another thermal. The additional component adds two additional free parameters: the spectral index or temperature
and another normalization. For all fits, the column density was fixed at the Galactic value, 5$\times$10$^{20}$ cm$^{-2}$ (Dickey \& Lockman, 1990)
and the redshift was fixed at 0.00928 (Ricardo et al. 2008)

\section{Results: Thermal Emission}

A single thermal component shows significant residuals in the PCA spectrum that are due to Cosmic 
X-ray background (CXB) fluctuations and additional source components (see Figure 1). A single thermal component with 
CXB fluctuations modeled is still poor fit to the spectrum, $\chi^{2}$ = 511 for 224 degrees of freedom, showing systematic residuals. The CXB spectrum in this energy range is relatively flat so that, compared to
a steep spectrum radio/X-ray source emission, it underpredicts the 2 - 7 keV emission and over predicts the 8 - 15 keV emission (see Figure 2).
An additional thermal component provides a significantly better with $\chi^{2}$ of 263 for 222 degrees of freedom (see Figure 3). The F statistic is
124 and indicates nearly 100\% significance of the additional component. The decrease in $\chi^{2}$ improves the chance probability from 0 to 3\%.
The temperature range for the two components are 0.87 - 0.88 and 2.0 - 3.3 keV and the abundance is 0.16 - 0.19 Solar. The abundance appears significantly lower than that reported from XMM, 0.31 - 0.77 Solar.
Gastaldello et al. (2007) analyzed quadrants within 7.5' for XMM while we measure averages over 30'.  Since the abundance decreases with radius, this discrepancy may be due to the radial abundance gradient.
NGC 5044 has a luminosity distance of 42.8 Mpc. The two thermal component fit gives 6.9$\times$10$^{42}$
ergs s$^{-1}$ for the cool component and 
9.8$\times$10$^{41}$
ergs s$^{-1}$ for the hotter one. Normalization and luminosity are given for all models in Table 2. All
luminosities are in the 0.5 - 15 keV band.
The two temperatures in the RXTE/ROSAT modeling are consistent with the radial two temperature analysis of
the Chandra and XMM observations (Buote et al. 2003). Their spatial analysis shows that in a two phase
model, the temperatures extend into the core with the higher component decreasing slightly with radius and
the cooler one rising slightly with radius.

\subsection{Results: Non-thermal Emission}

Addition of a powerlaw to the thermal and CXB emission provides the best fitting model.
The addition of a 
powerlaw with spectral index 2.64 - 2.82 and luminosity of 2.6$\times$10$^{42}$ ergs s$^{-1}$ provides the best fit 
to the data with a $\chi^{2}$ of 249 for 222 degrees of freedom (see Figure 4). 
Systematic residuals  that are visible in the single thermal model are eliminated or substantially reduced. 
The decrease in $\chi^{2}$ from
the thermal model improves the chance probability to 10\%. The reduced $\chi^{2}$ is higher than one due to random fluctuations in
a few PSPC channels and residuals from multiphase gas not fully modeled in the 3 - 6 keV. For example, ignoring PSPC bins 122 and 215 increases the probability to 25.3\% without significantly affecting the fit parameter
errors. The non-thermal component  is 32\% of the total
X-ray emission before point-source removal.

There is good agreement between the XMM abundance, 0.31 - 0.77 Solar, and that obtained with the best fitting, non-thermal PSPC/RXTE model, 0.26 - 0.47 Solar. 
Any discrepancies in abundance can not
account for the the non-thermal emission measured by the PCA since the non-thermal emission becomes dominant in the spectrum in the energy range 7 - 15 keV, above FeK line
complex.
%There are three possibilities for the hard X-ray component that was detected with either at the optical redshift or the X-ray redshift: %(1) a non-thermal component, 
%(2) a hot thermal component with small luminosity, (3) a slightly hotter component with a significant luminosity and the entire X-ray
%emitting gas having ~600 km s$^{-1}$ velocity with respect to the galaxies.

%For the second possibility, noting that the luminosity is 6.1$\times$10$^{40}$ ergs s$^{-1}$ and the temperature, 2.5 - 5.52 keV, it is %within the
%range of galactic hard sources in early-type galaxies (Matsumota et al. 1997).

%Both of the two component models are an improvement over the single component. The second thermal
%component model is considerable higher than that found by Buote et al and may be due to the sensitivity
%of the PCA to higher temperature gas. Equally likely is that the component is non-thermal.

%Need to compare the results to Buote I in my directory.

\subsection{Possible Contaminating Sources}

Seven discreet sources are identified in the ROSAT PSPC image that fall within the PCA field-of-view.
Spectra were extracted from each source and co-added for modeling. The background file used to
obtain the diffuse emission was also used in the point source analysis. For column density fixed at the Galactic value,
the best fit model for the point sources is thermal, 0.85 - 1.10 keV, with luminosity
9.5$\times$10$^{40}$ ergs s$^{-1}$ plus a powerlaw, $\alpha$ = 2.2, and luminosity 1.39$\times$10$^{41}$ ergs s$^{-1}$. The
90\% confidence upper limit for the non-thermal emission is 1.7$\times$10$^{41}$ ergs s$^{-1}$.
Both luminosities are given in the ROSAT energy band, 0.5 - 2.5 keV. Converting to the 0.5 - 15 keV energy band for comparison to the
non-thermal source emission gives a non-thermal point source emission of 3.1$\times$10$^{41}$ ergs s$^{-1}$ or 12\%
of the non-thermal emission. Using the upper limit on the non-thermal point source emission, the fraction extends up to 14\%. 
The diffuse, point source subtracted, non-thermal component is 2.2 - 3.0$\times$10$^{42}$ergs s$^{-1}$. Using the Chandra data, David et al. identified 8 point sources within the central region.
In the 0.1 - 80 kev band, these point sources have a combined luminosity of $<$5$\times$10$^{40}$ ergs s$^{-1}$. We used
the PSPC because its wider field of view better matches the PCA and our point source emission is an order of magnitude higher
and thus it is consistent with the Chandra modeling of the central region.

\section{Discussion}

\subsection{Non-thermal Characteristics of the Group Medium}

The NGC 5044
group has a velocity
relative to the microwave background of 2924 km s$^{-1}$ and a redshift of 0.009754 (0.0087, uncorrected) to the CMB
(H$_{0}$=73 km s$^{-1}$ Mpc$^{-1}$, $\Omega_{m}$=0.27, $\Omega_{v}$=0.73). This gives a
scale of 0.70 Mpc degree$^{-1}$. Ferguson \& Sandage (1990) modeled a data set consisting of 105
galaxies and found an optical core radius of 147 kpc and a central galaxy density
of 543 Mpc$^{-2}$. The high central density indicates that the group is quite compact and perhaps not yet virialized. These authors also report a relatively high
velocity dispersion, 474 km s$^{-1}$, though based on only 5 galaxies.  However, Cellone \& Buzzoni (2005), looking at dwarf galaxies near NGC 5044, find that
the group appears clearly defined in redshift space, with 
a mean heliocentric radial velocity, 
v$_{r}$ = 2461 +/- 84 km s$^{-1}$ 
(z = 0.0082), and a significantly more moderate dispersion, $\sigma$ = 290 km s$^{-1}$.

%Galactic extinction is 0.11 magnitudes based on HI (Burstein \& Heiles, 1982, AJ, 1165) and
%somewhat higher, 0.30, based on dust infrared emission (Schelegel et al., 1998, ApJ, 500, 525).

Early X-ray studies established that the intragroup gas is multiphase.  The cooler 
component in a two temperature component modeling is ( $\sim$ 0.7 keV) and the hot phase is
($\sim$1.4 keV), roughly what would be expected from gravitational heating from the central galaxy and group, respectively (Buote et al. 2003).
Morphological evidence of dynamical disturbance is visible in the high resolution X-ray image obtained with Chandra, including plumes, holes, compressions, and isophotal twisting 
in the core, suggest additional, non-gravitational forms of heat. Within the central 10 kpc, there are several radio quiet
cavities visible in the X-ray image that provide heating in the core (David et al. 2009). At larger radii, $\sim$50 kpc, 
the  sharp edge of a cold front is visible  (Gastaldello et al. 2009) indicative of heating by merger or interaction. 
The NGC 5044 group shows evidence of a recent dynamical encounter with a mass approximately 20\% of the group's virial mass (Gastaldello et al. 2009).

Recent GMRT imaging at low frequency (David et al. 2009) shows several diffuse radio source components extending out into the intragroup
medium: (1) emission at 235 MHz extending southwest approximately 51  kpc out from the nucleus of NGC 5044,  (2) 
emission at 235 MHz extending out approximately 75 kpc and detached from the nucleus to the southeast, and (3)
emission at 610 MHz along the same southeast axis much closer to the nucleus and not co-spatial with the 235 MHz emission. 

%These authors conclude that the lack of 610 MHz
%emission in the 235 MHz radio emission implies a steeper spectral index ($\alpha$ $\ge$1.6). compared to the 600 MHz emission ($\alpha$ $\le$1.6)
%which lacks 235 MHz emission. 
The lack of detection at 610 MHz in the detached source to the southeast, implies a steep spectral index, ($\alpha$ $\ge$1.6) (David et al. 2009).
Assuming the hard X-ray emission we detected with the PCA results from inverse
Compton (IC) emission from the same cosmic-ray electrons producing one of the group radio sources, our X-ray spectral index favors
the 235 MHz regions to the southeast for its origin. This region also has the highest radio flux at 235 MHz
(L. David 2009 Private Communication) and the largest volume.  Assuming that the radio sources have a nearly uniform
magnetic energy density implies that the detached southeastern source should be the primary source of IC hard X-rays since it has the largest
concentration of cosmic-ray electrons. This allows us to use the radio flux  and the X-ray flux
to eliminate the cosmic-ray energy distribution in calculating the average magnetic field. Using the expression for the
 inverse Compton flux in terms of the magnetic field (Henriksen 1999) gives an average projected magnetic field of 0.034 $\mu$Gauss. 

The energy loss rates for cosmic ray electrons via inverse-Compton cooling by cosmic microwave (CMB) photons compared to 
synchrotron via a magnetic field is equal
for a field of about 3 $\mu$G. The average field for this detached (from the AGN) source is a factor of $\sim$100 lower
meaning that IC cooling will dominate. The inverse-Compton cooling time is 2.4$\times$10$^{6}$years/($\gamma_{min}$/1000).

Equations 1 - 3 in Henriksen (1999) are used to calculate the 
cosmic ray electron energy density from the observed inverse-Compton flux and the
cosmic ray volume inferred from of the radio emitting region. The detached source is estimated to be
described by a sphere with a diameter of 3 arc minutes using the radio image. This gives
6.05$\times$10$^{68}$ cm$^{3}$ for the co-spatial volume of cosmic-ray electrons and magnetic field. 
We calculate a cosmic ray energy density of 3.6$\times$10$^{-12}$ ergs cm$^{-3}$. The volume of cosmic-rays,
V$_{cr}$, could be larger than the observed radio lobes with the magnetic field filling only a fraction (f$_{m}$) (=V$_{m}$/V$_{cr}$) of the
true cosmic-ray volume. In this case, the radio source would only occupy 
V$_{cr}$(f$_{m}$) = 6.05$\times$10$^{68}$ cm$^{3}$ and the true cosmic ray volume would be perhaps significantly larger. 
For an observed flux of non-thermal X-rays, increasing the cosmic ray volume decreases the cosmic-ray energy density. Alternatively,
defining (f$_{cr}$) as the filling factor of cosmic-rays, Fabian et al. (2002) proposed that certain geometries (e.g., a shell of cosmic-rays), would decrease the calculated
volume and increase the cosmic-ray energy density. 
The magnetic field energy density
is calculated to be 1.45$\times$10$^{-16}$ ergs cm$^{-3}$. The equipartition magnetic field, $\sim$12 $\mu$G, is 353 times higher than the average field so that the
magnetic field and cosmic-ray electrons are significantly out of equipartition. A very small magnetic filling factor, f$_{m}$$<$0.01\%, would
be required for equipartition. We calculate an average gas energy density, calculated at the center of the source, of 6.9$\times$10$^{-11}$
ergs cm$^{-3}$ for an electron density of 10$^{-3}$ cm$^{-3}$ and temperature of 5$\times$10$^{7}$ K (Buote et al. 2003). 
A magnetic filling factor of less than one is perhaps undesirable in that it would exacerbate the lack of energy balance between 
cosmic-ray energy density and gas energy density and would increase the dynamical instability of the cosmic-ray region.
For the radio lobe with f$_{m}$ = 1 to be in equilibrium with
its surrounding intracluster gas, the energy density from sources other than the cosmic-ray electrons that contribute
to the non-thermal radiation, must be an order of magnitude higher than the electron
cosmic-ray energy density. This is given by the
K parameter (Fabian et al. 2002; Dunn \& Fabian 2004), the ratio of energy density from the other sources to cosmic-ray electrons, and includes protons
and sub GeV cosmic-ray electrons. We infer
that K must be approximately 12 to obtain equilibrium. For the simplest case, K =1,  the surrounding ICM is compressing the detached source.
Since both K and the cosmic-ray filling factor, f$_{cr}$, $<$1 both serve to increase the cosmic-ray energy density, the ratio, K/f$_{cr}$, must be $\sim$19
to have energy balance between the non-thermal and thermal sources at the radio lobe. 

\subsection{Largest Radio Source is a Merger Remnant}

The cosmic-ray energy density is average for the detached source compared with radio lobes (compilation in Isobe et al. 2009), however, the magnetic field
density is much lower with only 3C 403 and 3C 452 comparable (Croston et al. 2005). 
Thus the detached radio source is significantly out of equipartition with an U$_{cr}$/U$_{B}$ of $\sim$2.5$\times$10$^{4}$. Radio lobes are typically
in equipartition or slightly electron dominated, with lower energy density ratio, between 1 and 10 (comparison sample in Croston et al. 2005).
Correlation between both cosmic-ray energy density and magnetic field
energy density with total lobe size (Isobe et al. 2009) shows a general decrease with increasing radio lobe
size for each. This can generally be understood by the expansion of the lobe against the surrounding gas until it reaches
equilibrium between gas pressure and non-thermal pressure inside the lobe. The largest lobes then have the lowest non-thermal densities.
The highest non-thermal densities are the smallest lobes, yet to fully expand. Thus, the lobes should expand
only until equilibrium is reached and a small non-thermal to thermal energy density as we find for the detached lobe is difficult to explain
in this context. Comparison of the detached source with 
the radio lobes shows that it's diameter of 35 kpc is much smaller than the radio lobe sample range, 50 - 500 kpc. It clearly
does not follow the correlation in that extrapolation of the figure to its size would predict a cosmic-ray electron energy density of
approximately 10$^{-9}$ ergs cm$^{-3}$. The observed value for the detached source is $\sim$200 below the predicted value. Its predicted magnetic field
energy density is 30,000 times too small. This indicates that the southeastern radio source is not likely a detached lobe that was once
connected to the AGN, but is a relic of a merger event. 

The energetics of formation via merger shock is plausible. Because the inverse-Compton cooling time is short, $\le$10$^{7}$ years, continuous
re-acceleration must take place. For a nominal lifetime of 10 billion years, the relic would require require $\sim$3.3$\times$10$^{60}$ ergs of energy input. Gastaldello et al. (2007) calculate a virial mass of 3.7$\times$10$^{13}$ Solar Masses within 860 kpc for the NGC 5044 group.  They argue for a merger with a subclump with 20\% of the virial mass based on morphological similarities between identified cold fronts in the X-ray 
data and merger simulations (Ascasibar \& Markevitch 2006). Their supporting evidence for a merger includes presence
of a distinct sub-group 1.4 Mpc away and a relatively high velocity of the cD galaxy to the mean, 150 km s$^{-1}$ (Mendel et al. 2007).
 The fraction of energy crossing the shock front that goes into cosmic-rays can be estimated based using these masses
 and simulations. The kinetic energy from the merger is $\sim\rho$V$_{flow}^{3}$R$_{cl}^{2}$ (Miniati et al. 2000).
The flow velocity is the free-fall velocity, 620 km s$^{-1}$ for the mass of the group, 3.7x10$^{13}$ Solar Masses within 860 kpc.
The adiabatic sound speed is 520 km s$^{-1}$ using our emission weighted temperature, 1.04 keV. This gives 
 1.8$\times$10$^{44}$ ergs s$^{-1}$ of kinetic energy through the shock. The fractional conversion to cosmic-rays for a weak shock, M$\sim$1.2, is $<$5\% (Kang, Jones \& Geiseler 2002); nearly equal to the combined inverse-Compton and synchrotron component luminosities.
 
 \section{Conclusions}

Fitting RXTE PCA and a co-spatial ROSAT PSPC spectrum indicates that the preferred model for the X-ray emission includes
non-thermal emission that is accurately modeled with a power law with a steep spectral index, 2.6 - 2.8, and
X-ray luminosity of 2.6$\times$10$^{42}$ ergs s$^{-1}$ in the 0.5 - 15 keV band. After accounting for non-thermal emission
from bright point sources in the PSPC, the non-thermal luminosity is reduced to 2.3$\times$10$^{42}$ ergs s$^{-1}$ or 28\% of
the total X-ray emission. The diffuse, point source subtracted, non-thermal component is 2.2 - 3.0$\times$10$^{42}$ergs s$^{-1}$, with 90\% confidence. We find a cosmic-ray electron 
energy density,  3.6$\times$10$^{-12}$,  and average magnetic field, 0.034$\mu$Gauss,  in the largest  radio emitting region
within the group. The ratio of
cosmic-ray electron energy density to magnetic field energy density, $\sim$2.5$\times$10$^{4}$,  is much larger than is typical
of radio lobes, mainly due to the low magnetic field. The relic's small size and low non-thermal energy density,
 contradicts the size-energy relationship found for radio lobes further suggesting that the relic is formed by non-AGN process.
 Calculation of the energy in cosmic-rays accelerated by 
merger shocks is consistent with the non-thermal luminosity arguing that the relic is the result of the recent merger
in the NGC 5044 group.

I acknowledge support from NASA for this project through grant NNX07AG34G.

%\plotone{pspc_on_dssred.ps}

%\includegraphics{model_comp.ps}
% \includegraphics[width=60mm]{myfig.eps}
  %\includegraphics[height=60mm]{myfig.eps}
  %\includegraphics[scale=0.75]{myfig.eps}
%\includegraphics[angle=-90,width=160mm]{model_comp.ps}
%Figure 1: The best model is: thermal (dashed), powerlaw (dash-dot), and total (solid).

%\includegraphics[angle=-90,width=160mm]{1rs.ps}
%Figure 1: A single thermal component fit to the combined PSPC and PCA spectra. Significant
%residual hard emission due to shocked gas, non-thermal emission, contaminating AGN, or hard
%X-ray background  (HXB) fluctuations are indicated.

\includegraphics[angle=-90,width=160mm]{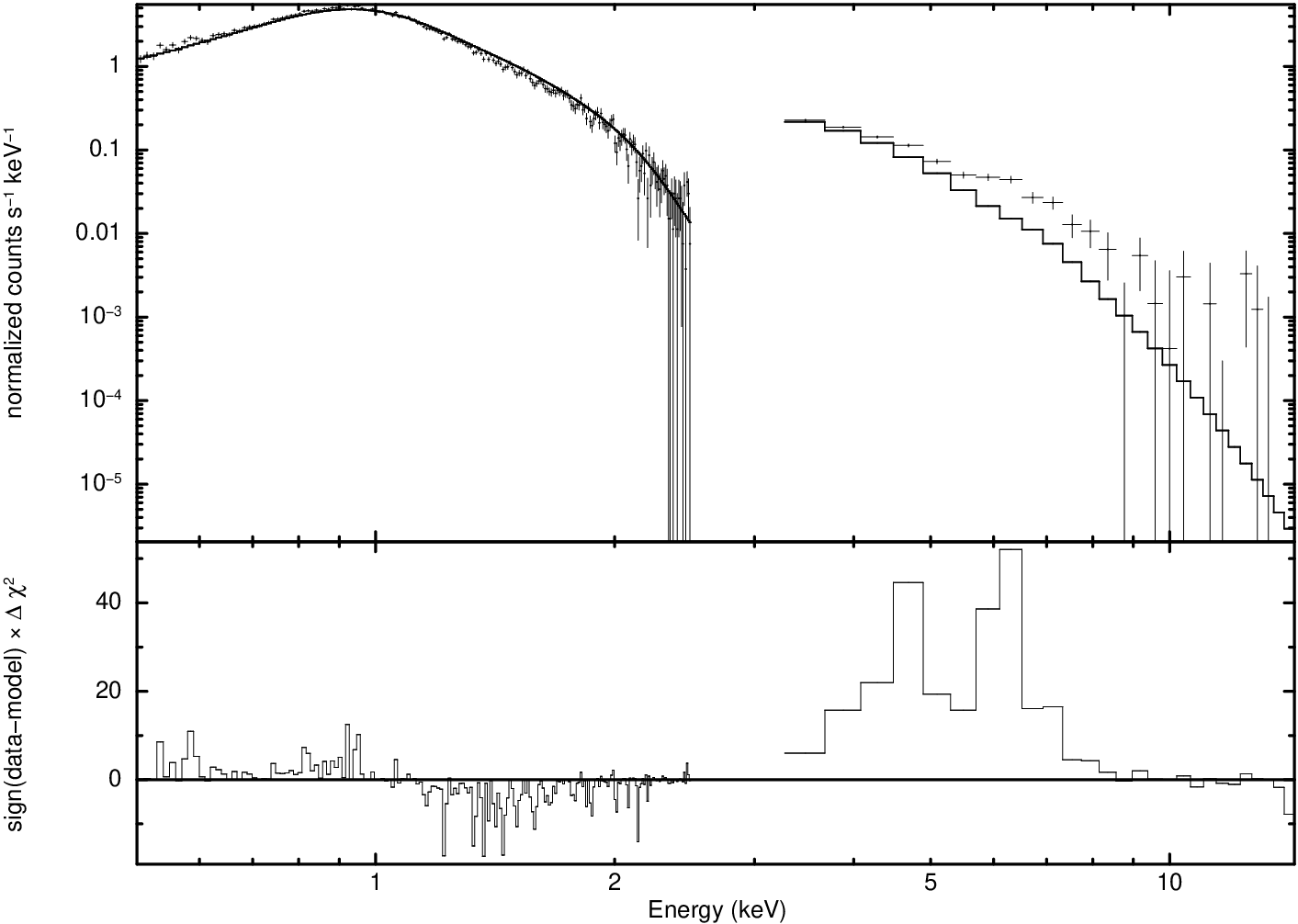}
 Figure 1: A single thermal component shows significant residuals in both the PSPC and PCA spectrum that are due to Cosmic 
X-ray background (CXB) fluctuations and additional source components. 

\includegraphics[angle=-90,width=160mm]{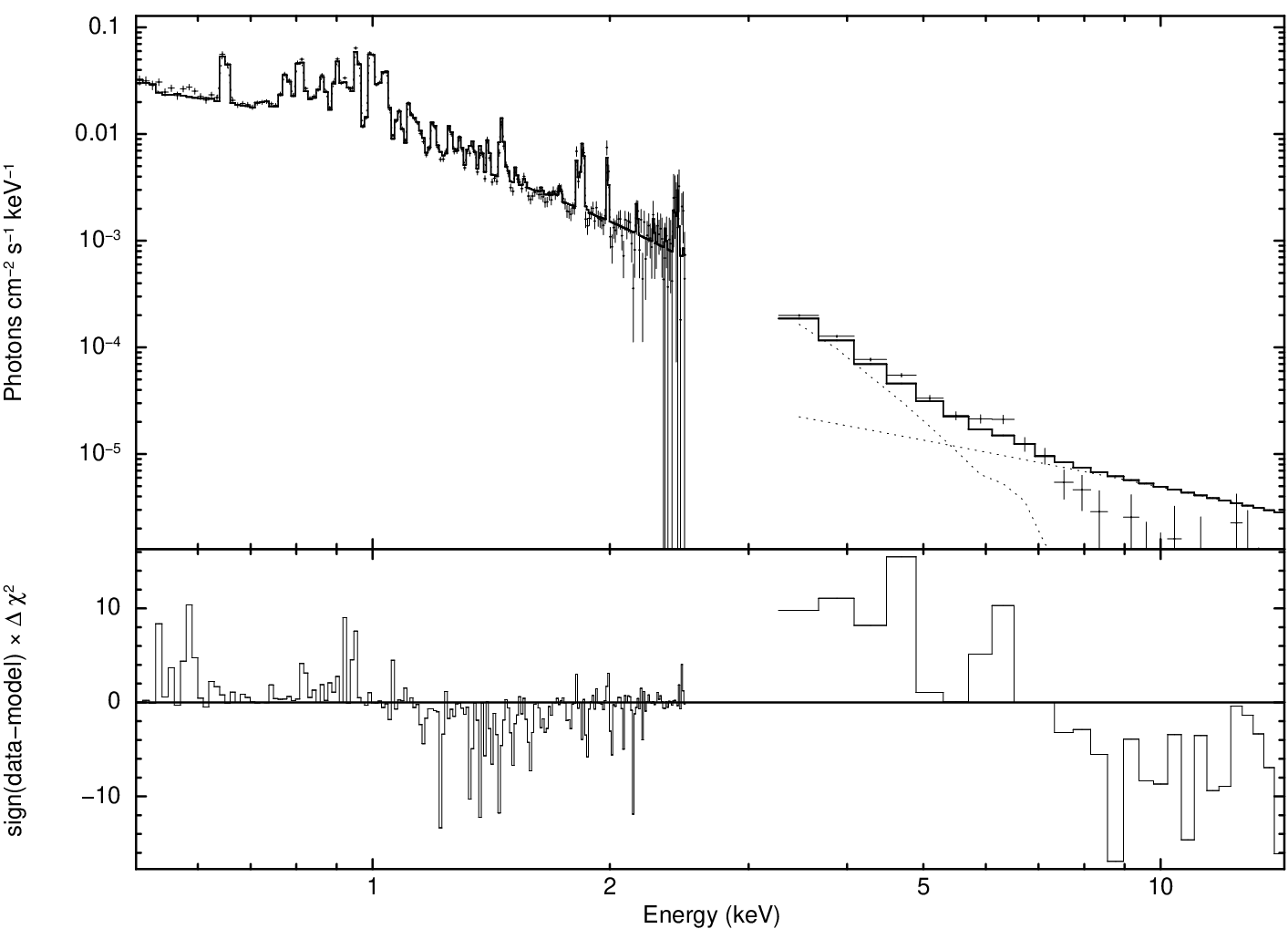}
 Figure 2: A single thermal component with hard Cosmic X-ray background (CXB) fluctuations is a poor fit to the spectrum. The spectrum is relatively flat so that, compared to a steep spectrum X-ray source, it underpredicts the 2 - 7 keV emission and over predicts the 8 - 15 keV emission. In figures 2 -4, the dependence of the unfolded spectrum on the assumed model is visible.

\includegraphics[angle=-90,width=160mm]{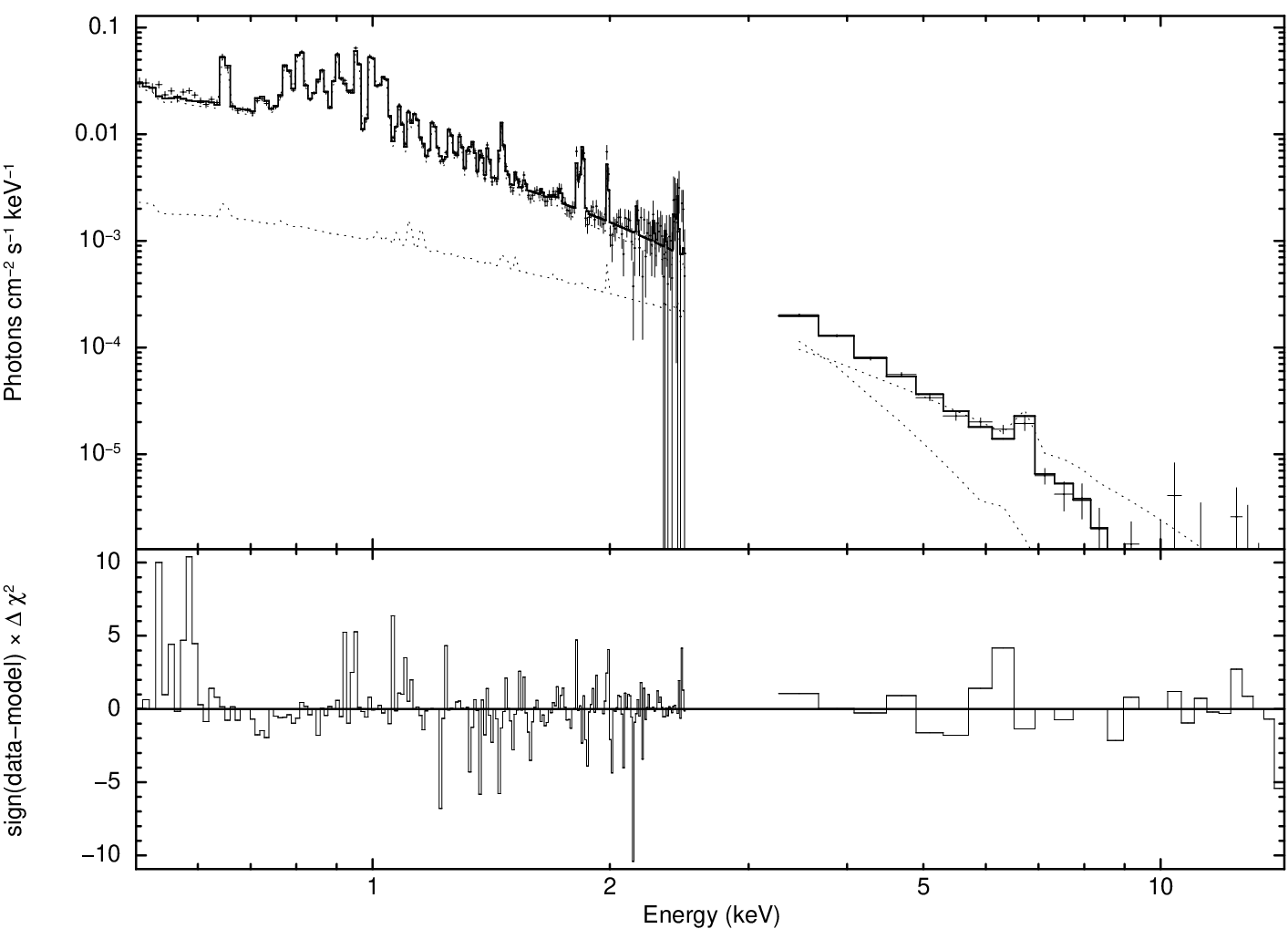}
 Figure 3: Addition of a second thermal component provides a better fit. However, systematic residuals remain in the PSPC below 1 keV.

\includegraphics[angle=-90,width=160mm]{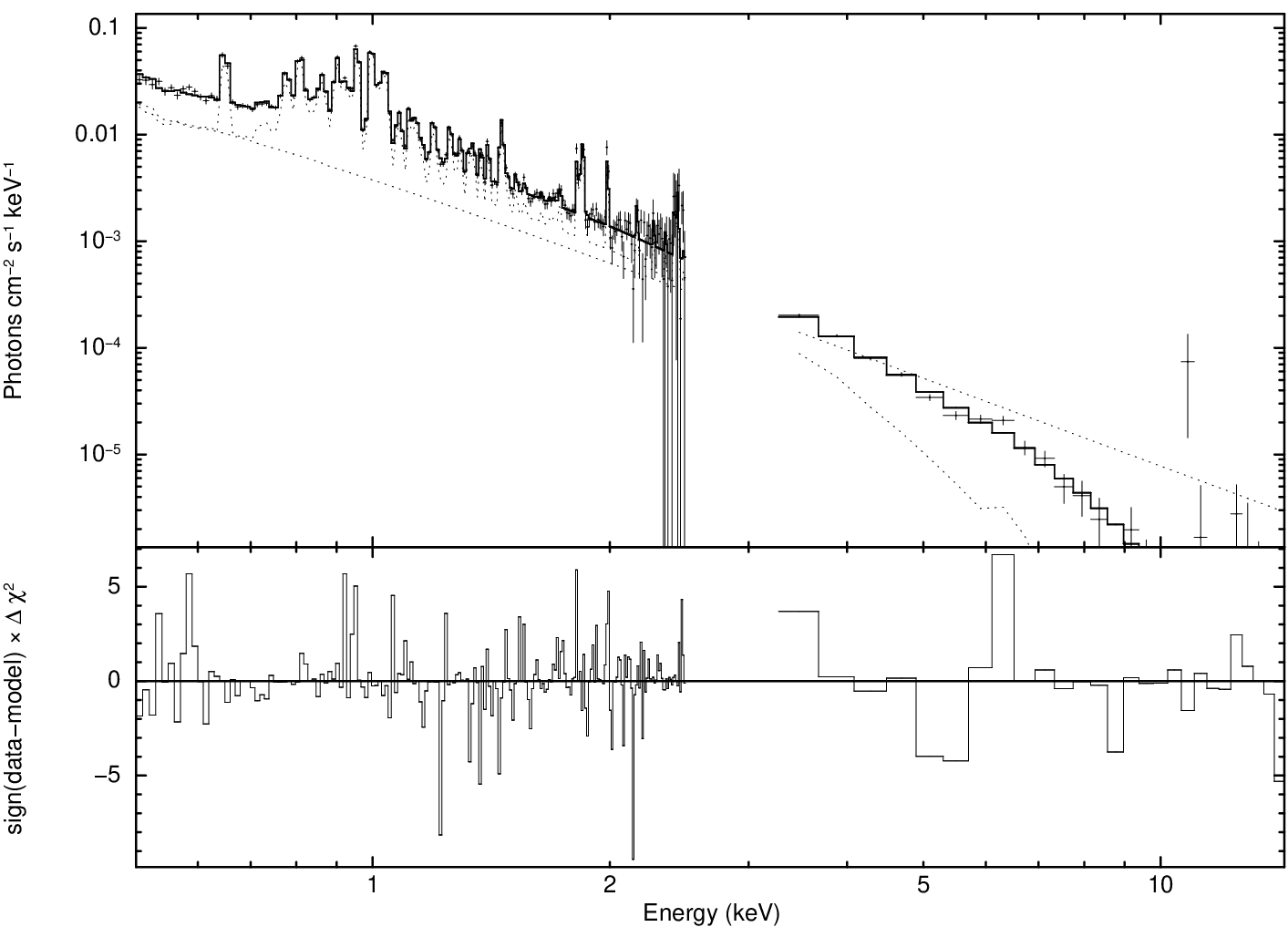}
 Figure 4: Addition of a powerlaw to the thermal and CXB emission provides the best fitting model. All residuals are random.

%\includegraphics[angle=-90,width=160mm]{fig2.ps}
% Figure 1: Light curve is clipped at 30 counts per 100 second bin to filter out spikes.
 
%\includegraphics[angle=-90,width=160mm]{fig3.ps} 
%Figure 3: Systematic residuals in the PCA at E$>$5 keV for the thermal model.

%\includegraphics[angle=-90,width=160mm]{fig4.ps} 
%Figure 4: Addition of a powerlaw significantly diminishes the residual emission.

%\plotone{pspc_on_dssred.ps}

%\plotone{acis_on_dss_and_nvss_on_acis_new.ps}

%\plotone{nvss_on_dss.ps}

\vfil\eject

%\plotone{ros_point_image.ps}

\begin{deluxetable}{lccccc}
\tabletypesize{\scriptsize}
\tablewidth{0pt}
\tablecolumns{8}
\tablecaption{NGC 5044 Group  Spectral Fits \label{tbl-1}}
\tablehead{\colhead{Model} &\colhead{kT$_{1}$ (keV)} & \colhead{Abundance} &
 \colhead{$\Gamma_{X}$} & \colhead{kT$_{2}$ (keV)} & \colhead{$\chi^2$/dof}}
\startdata
Thermal + CXB  & - & - & - & - & 511/224 \\
Thermal + Powerlaw + CXB &  0.89 - 0.93 & 0.26 - 0.47 & 2.64 - 2.82 & - &  249/222 \\
Thermal + Thermal + CXB &  0.87 - 0.88  &  0.16 - 0.19 & - & 2.36 - 3.63 & 263/222 \\
\enddata
\end{deluxetable}

\begin{deluxetable}{lcccccc}
\tabletypesize{\scriptsize}
\tablewidth{0pt}
\tablecolumns{8}
\tablecaption{NGC 5044 Group Normalization, Flux, and Luminosity \label{tbl-2}}
\tablehead{\colhead{Model} &\colhead{Normalization(1)} 
& \colhead{Normalization(2)}  & \colhead{Flux(1)} &   \colhead{Flux(2)} & \colhead{L$_{x}$(1)} & \colhead{L$_{x}$(2)}}
\startdata 
Thermal(1) + Powerlaw(2) &  0.026 - 0.040 & 0.0036 - 0.0049 &  2.68e-11 & 1.24e-11 & 5.5e42  & 2.6e42\\ 
Thermal(1) +  Thermal(2) & 0.053 - 0.060 & 0.004 - 0.006 & 3.28e-11 & 4.73e-12 & 6.9e42 & 9.8e41 \\
\enddata
\end{deluxetable}

\end{document}